\documentclass[aps,prc,preprint,superscriptaddress,showpacs]{revtex4}
\usepackage[dvips]{graphicx}
\usepackage{amsmath,amssymb}
\usepackage{ulem}

\newcommand{\beq}{\begin{equation}}
\newcommand{\eeq}{\end{equation}}
\newcommand{\bea}{\begin{eqnarray}}
\newcommand{\eea}{\end{eqnarray}}

\DeclareSymbolFont{boldletters}{OML}{cmm} {b}{it}
\DeclareSymbolFontAlphabet{\mathbit}{boldletters}
\DeclareMathSymbol{\alpha}{\mathalpha}{letters}{"0B}
\DeclareMathSymbol{\beta}{\mathalpha}{letters}{"0C}
\DeclareMathSymbol{\gamma}{\mathalpha}{letters}{"0D}
\DeclareMathSymbol{\delta}{\mathalpha}{letters}{"0E}
\DeclareMathSymbol{\epsilon}{\mathalpha}{letters}{"0F}
\DeclareMathSymbol{\zeta}{\mathalpha}{letters}{"10}
\DeclareMathSymbol{\eta}{\mathalpha}{letters}{"11}
\DeclareMathSymbol{\theta}{\mathalpha}{letters}{"12}
\DeclareMathSymbol{\iota}{\mathalpha}{letters}{"13}
\DeclareMathSymbol{\kappa}{\mathalpha}{letters}{"14}
\DeclareMathSymbol{\lambda}{\mathalpha}{letters}{"15}
\DeclareMathSymbol{\mu}{\mathalpha}{letters}{"16}
\DeclareMathSymbol{\nu}{\mathalpha}{letters}{"17}
\DeclareMathSymbol{\xi}{\mathalpha}{letters}{"18}
\DeclareMathSymbol{\pi}{\mathalpha}{letters}{"19}
\DeclareMathSymbol{\rho}{\mathalpha}{letters}{"1A}
\DeclareMathSymbol{\sigma}{\mathalpha}{letters}{"1B}
\DeclareMathSymbol{\tau}{\mathalpha}{letters}{"1C}
\DeclareMathSymbol{\upsilon}{\mathalpha}{letters}{"1D}
\DeclareMathSymbol{\phi}{\mathalpha}{letters}{"1E}
\DeclareMathSymbol{\chi}{\mathalpha}{letters}{"1F}
\DeclareMathSymbol{\psi}{\mathalpha}{letters}{"20}
\DeclareMathSymbol{\omega}{\mathalpha}{letters}{"21}
\DeclareMathSymbol{\varepsilon}{\mathalpha}{letters}{"22}
\DeclareMathSymbol{\vartheta}{\mathalpha}{letters}{"23}
\DeclareMathSymbol{\varpi}{\mathalpha}{letters}{"24}
\DeclareMathSymbol{\varrho}{\mathalpha}{letters}{"25}
\DeclareMathSymbol{\varsigma}{\mathalpha}{letters}{"26}
\DeclareMathSymbol{\varphi}{\mathalpha}{letters}{"27}
\DeclareMathSymbol{\Gamma}{\mathalpha}{letters}{"00}
\DeclareMathSymbol{\Delta}{\mathalpha}{letters}{"01}
\DeclareMathSymbol{\Theta}{\mathalpha}{letters}{"02}
\DeclareMathSymbol{\Lambda}{\mathalpha}{letters}{"03}
\DeclareMathSymbol{\Xi}{\mathalpha}{letters}{"04}
\DeclareMathSymbol{\Pi}{\mathalpha}{letters}{"05}
\DeclareMathSymbol{\Sigma}{\mathalpha}{letters}{"06}
\DeclareMathSymbol{\Upsilon}{\mathalpha}{letters}{"07}
\DeclareMathSymbol{\Phi}{\mathalpha}{letters}{"08}
\DeclareMathSymbol{\Psi}{\mathalpha}{letters}{"09}
\DeclareMathSymbol{\Omega}{\mathalpha}{letters}{"0A}
\newcommand{\mbit}[1]{{\mathbit#1}}

\newcommand{\dsl}[2]{{#1}{\mbox{\hspace{-8pt}\hspace{#2}$\not$}
\hspace{8pt}\hspace{-#2}}} 

\newcommand{\dslpar}{\dsl{\partial}{1pt}}

\begin{document}
\title{Meson loop effect on high density chiral phase transition}

\author{Tomohiko Sakaguchi}
\affiliation{Department of Physics, Graduate School of Sciences, Kyushu University,
             Fukuoka 812-8581, Japan}

\author{Kouji Kashiwa}
\affiliation{Department of Physics, Graduate School of Sciences, Kyushu University,
             Fukuoka 812-8581, Japan}

\author{Masayuki Matsuzaki}
\email[]{matsuza@fukuoka-edu.ac.jp}
\affiliation{Department of Physics, Graduate School of Sciences, Kyushu University,
             Fukuoka 812-8581, Japan}
\affiliation{Department of Physics, Fukuoka University of Education, 
             Munakata, Fukuoka 811-4192, Japan}

\author{\\Hiroaki Kouno}
\affiliation{Department of Physics, Saga University,
             Saga 840-8502, Japan}

\author{Masanobu Yahiro}
\affiliation{Department of Physics, Graduate School of Sciences, Kyushu University,
             Fukuoka 812-8581, Japan}

\date{\today}

\begin{abstract}
We test the stability of the mean-field solution in 
the Nambu--Jona-Lasinio model. For stable solutions with respect to 
both the $\sigma$ and $\pi$ directions, we investigate effects of 
the mesonic loop corrections of $1/N_c$, which correspond to 
the next-to-leading order in the $1/N_c$ expansion, 
on the high density chiral phase transition. 
The corrections weaken the first order phase transition and shift the critical 
chemical potential to a lower value. At $N_c=3$, however, instability of the 
mean field effective potential prevents us from determining the minimum of the 
corrected one. 
\end{abstract}

\maketitle

At high temperature and density quantum chromodynamics (QCD) 
undergoes qualitative changes of great physical interest. Although many works 
are done, aspects of the transition region are not 
under full theoretical control.
It is mandatory to deepen its understanding 
in order to understand the structure of compact stars and 
the history of the early 
universe, as well as results of ultra relativistic heavy ion experiments. 
With the aid of the progress in computer power, lattice simulations 
have become feasible for thermal system with zero or small density. 
In general, chiral restoration and deconfinement are expected to occur 
simultaneously~\cite{Kog}. One of the most important recent findings is 
strong correlations in the deconfined quark gluon plasma just 
above the critical temperature; on the one hand it appears as the near 
perfect fluidity~\cite{Lee,GM,Shu} and on the other hand as the mesonic 
correlations~\cite{AH,Gho}. 

As an approach complementary to the first-principle lattice QCD simulation, 
we can consider effective models. In particular, they are even indispensable 
at high density where lattice QCD is not applicable due to the sign 
problem. One of them is the Nambu--Jona-Lasinio (NJL) model. Since it was 
proposed~\cite{NJ1,NJ2}, although this is a model of chiral symmetry that does 
not possess a confinement mechanism, this model has been widely 
used~\cite{Kle,HK1} in the mean field approximation, for example, for 
analyses of the critical end point of chiral transition on the temperature 
($T$) - chemical potential ($\mu$) plane~\cite{AY,BR,Sca,Fuj}. 
Only a few studies that go beyond the 
mean field 
approximation were reported. 
Nikolov et al.~\cite{Nik} extended the mean field approximation 
to next-to-leading order 
in $1/N_c$ which includes the meson-loop contributions. 
The extension preserves the symmetry property of 
the theory. They studied the meson loop effect in the case of $T=\mu=0$ 
and concluded its importance due to the light pion mass. 
H\"ufner et al.~\cite{Huf} formulated thermodynamics of the NJL model 
to order $1/N_c$ and 
Zhuang et al.~\cite{Zhu} presented numerical results 
by using the formulation. Their analyses are mainly made 
for thermal system in the limit of zero $\mu$. Although 
only a result is reported for the case that both $T$ and $\mu$ are finite, 
one can expect from the result that the meson loop corrections are 
important also for finite $\mu$.

 In this Letter our attention is focused on the finite-density chiral phase 
transition in the limit of zero $T$. 
We first investigate the stability of the mean field solution and for the stable mean field solution we further evaluate the magnitude of corrections of 
meson loops, that is, of the next-to-leading order in the $1/N_c$ expansion, 
using the auxiliary field method of Kashiwa and Sakaguchi~\cite{KS} which 
is essentially equivalent to the formulation of H\"ufner et al.~\cite{Huf}. 

The Lagrangian density of the NJL model is
\begin{eqnarray}
 \mathcal{L} = \bar{q}(x)(i \dslpar - m_0) q (x)
  +\frac{G}{2 N_c} ( (\bar{q}(x) q (x))^2 
+ (\bar{q}(x) i \gamma_5 \tau_a q (x) )^2 ) \ , 
\end{eqnarray}
where 
$q(x)$ stands for two flavor quarks, 
and $N_c$ for the number of color degrees of freedom.
$m_0$ is the current quark mass and $\tau_a$ is the Pauli matrix.
$G$ is the coupling constant of a four-Fermi interactions
scaled by the color number $N_c$ in order to use the $1/N_c$
expansion later. 
The partition function for the NJL model reads
\begin{eqnarray}
 Z &=& \int \mathcal{D} q \mathcal{D} \bar{q}
\exp \Bigg[ - \int_{\beta} d^4 x
\bigg(  \bar{q} (x)
(\partial_{\mu} \gamma^{\mu}+m_0 - \mu \gamma_4 
-J_{\sigma}(x)-i \gamma_5 \vec{\tau} \cdot \vec{J_{\pi}}(x)) q(x) \nonumber \\
&&-\frac{G}{2 N_c}  
\left( (\bar{q}(x) q(x))^2 + (\bar{q}(x) i \gamma_5 \tau_a q(x) )^2 \right) 
-\frac{N_c}{2 G}  ( J_{\sigma}^2(x) + {\vec{J}_{\pi}}^{\,2}(x) ) 
\bigg)    \Bigg] \ . \label{partition}
\end{eqnarray}
Here we introduced external fields $J_{\sigma}$ and $\vec{J}_{\pi}$. 
The last term in Eq.~(\ref{partition}) is introduced for later convenience. 
$\mu$ stands for the chemical potential of quarks.

The partition function~(\ref{partition}) has the four-Fermi interactions, 
accordingly we cannot carry out the Gaussian integration with respect 
to the quarks. Therefore,
we introduce the auxiliary fields $\sigma(x)$ and $\vec{\pi}(x)$ 
by using of the identity
\begin{eqnarray}
 1 = \int \mathcal{D} \sigma 
  \exp \left[- \frac{N_c}{2 G} \int_{\beta}
	d^4 x \left(\sigma(x) + \frac{G}{N_c} \bar{q}(x)q(x)   \right)^2
       \right] \ , 
\end{eqnarray}
and the corresponding one for $\vec\pi$. 
After integrating out quarks and shifting the auxiliary fields 
$(\sigma (x) - J_{\sigma} (x)) \mapsto \sigma (x)$ and 
$(\vec{\pi} (x) - \vec{J}_{\pi} (x)) \mapsto \vec{\pi} (x)$, 
we obtain the partition function of the auxiliary fields,
\begin{eqnarray}
 Z = \int \mathcal{D} \sigma \mathcal{D} \vec{\pi}
\exp \left[ - N_c I[\sigma, \vec{\pi} ]   \right] \ , \label{auxiliarypartition}\end{eqnarray}
with 
\begin{eqnarray}
 I[\sigma, \vec{\pi}] &=& \int_{\beta} d^4 x 
\left(\frac{1}{2G} \left( \sigma^2(x) + \vec{\pi}^2(x) \right)
+ \frac{1}{G} \left( \sigma (x) J_{\sigma}(x) 
+ \vec{\pi}(x) \cdot \vec{J}_{\pi}(x) \right)  \right) \nonumber \\
\noalign{\vspace{5mm}}
&&- \ln \det \left( \partial_{\mu} \gamma_{\mu} + m_0 - \mu \gamma_4 
+ \sigma (x) + i \gamma_5 \vec{\tau} \cdot \vec{\pi} (x) \right) \ .
\label{auxiliaryaction}
\end{eqnarray}

The generating function for the connected Green's function $W$ 
is defined as
\begin{eqnarray}
 Z[J_{\sigma}, \vec{J}_{\pi}] \equiv 
\exp \left[-N_c W[J_{\sigma}, \vec{J}_{\pi}]\right] \ .
\label{Wdefinition}
\end{eqnarray}
The fields $\varphi_{\sigma}$ and $\vec{\varphi}_{\pi}$ are also defined as 
\begin{eqnarray}
\frac{\varphi_{\sigma}(x)}{G} \equiv 
\frac{\delta W}{\delta J_{\sigma}(x)} = \frac{\langle \sigma (x) \rangle}{G}
=- \frac{1}{N_c} \langle \bar{q} (x)q (x) \rangle - \frac{J_{\sigma}(x)}{G} \ 
\label{fields}
\end{eqnarray}
for $\varphi_{\sigma}$ 
and the corresponding one for $\vec{\varphi}_{\pi}$. 
The fields 
represent the vacuum expectation values of the auxiliary fields 
in the external fields $J_{\sigma} (x)$ and 
$\vec{J}_{\pi} (x)$. 
The effective action is defined by the Legendre transformation
\begin{eqnarray}
 \Gamma \left[\varphi_{\sigma}, \vec{\varphi}_{\pi} \right]
= W [ J_{\sigma}, \vec{J}_{\pi} ] 
- \frac{1}{G} \int_{\beta} d^4 x 
\left( \varphi_{\sigma}(x) J_{\sigma}(x) 
+ \vec{\varphi}_{\pi}(x) \cdot \vec{J}_{\pi}(x) \right) \ .
\end{eqnarray}
Setting $J_{\sigma}(x)$ and $\vec{J}_{\pi}(x)$ to 
constants leads to the effective potential 
\begin{eqnarray}
 \Gamma \left[\varphi_{\sigma}, \vec{\varphi}_{\pi} \right]
\stackrel{J_{\sigma}, \vec{J}_{\pi} \mapsto
\mathrm{const.}}{\Longrightarrow}
\beta V \mathcal{V} (\bar{\varphi}_{\sigma}, \vec{\bar{\varphi}}_{\pi}) \ , 
\end{eqnarray}
where note that $\varphi_{\sigma} (x)$ and $\vec{\varphi}_{\pi} (x)$ 
become constants 
$\bar{\varphi}_{\sigma}$ and $\vec{\bar{\varphi}}_{\pi}$, respectively.

In order to carry out the path integral 
in Eq. (\ref{auxiliarypartition}), we expand 
Eq.~(\ref{auxiliaryaction}) 
around the classical solution $(\sigma_0, \vec{\pi}_0)$: 
\begin{eqnarray}
I = I_0 + \frac{1}{2} \int_{\beta} d^4 x d^4 y 
(\sigma-\sigma_0 \,\, \vec{\pi}-\vec{\pi}_0)
\left(
\begin{array}{ll}
I_{\sigma \sigma}^{(2)} & I_{\sigma \pi}^{(2)} \\
I_{\pi \sigma}^{(2)} & I_{\pi \sigma}^{(2)} 
\end{array}
\right)
\left(
\begin{array}{l}
\sigma-\sigma_0 \\
\vec{\pi}-\vec{\pi}_0
\end{array}
\right) + \cdots \ ,
\label{Iexpansion}
\end{eqnarray}
\begin{eqnarray}
 I_0 &=& \int_{\beta} d^4 x \left(\frac{1}{2 G} 
(\sigma_0^2 + \vec{\pi}_0^2) 
+ \frac{1}{G} (\sigma_0 J_{\sigma} + \vec{\pi}_0 \cdot \vec{J}_{\pi}) 
\right) \\
&-& \ln \det (\gamma_{\mu} \partial_{\mu} + m_0 -\mu \gamma_4 + \sigma_0 
+ i \gamma_5 \vec{\tau} \cdot \vec{\pi}_0) \ , 
\nonumber
\end{eqnarray}
\begin{eqnarray}
 I_{\hat{a} \hat{b}}^{(2)} = \frac{1}{G} \delta^{(4)}(x-y) 
\delta_{\hat{a} \hat{b}} + \mathrm{tr} [
\Gamma_{\hat{a}} S(x,y:\sigma_0,\vec{\pi}_0)
\Gamma_{\hat{b}} S(y,x:\sigma_0,\vec{\pi}_0) ] \ ,
\end{eqnarray}
where $\Gamma_{\hat{a}} =1$ for $\hat{a} = \sigma$ and  $\gamma_5 \tau_a$ for 
$\hat{a} = \pi_a \,\,\,\, (a = 1, 2, 3)$.
Here the classical solutions are governed by 
\begin{eqnarray}
&& \left. \frac{\delta I}{\delta \sigma (x)} 
\right|_{\sigma=\sigma_0, \vec{\pi} = \vec{\pi}_0} 
= \frac{1}{G}(\sigma_0(x)+J_{\sigma}(x)) -
\mathrm{tr}S(x,x:\sigma_0(x),\vec{\pi}_0(x)) 
= 0 \ ,  \\
&& \left. \frac{\delta I}{\delta \vec{\pi} (x)} 
\right|_{\sigma=\sigma_0, \vec{\pi} = \vec{\pi}_0} 
= \frac{1}{G}(\vec{\pi}_0(x)+\vec{J}_{\pi}(x)) -
\mathrm{tr} \left[i \gamma_5 \vec{\tau} 
S(x,x:\sigma_0(x),\vec{\pi}_0(x)) \right] 
= 0 \ ,
\end{eqnarray}
\begin{eqnarray}
 (\gamma_{\mu} \partial_{\mu} + m_0 -\mu \gamma_4 + \sigma_0(x) 
+ i \gamma_5 \vec{\tau} \cdot \vec{\pi}_0(x)) 
S(x,y:\sigma_0(x),\vec{\pi}_0(x)) = \delta^{(4)}(x-y) \ ,
\end{eqnarray}
where $S(x,y:\sigma_0,\vec{\pi}_0)$ denotes the quark propagator in the
external fields $J_{\sigma}(x)$. 
Making change of variables such that 
$(\sigma (x) - \sigma_0 (x)) \mapsto \sigma (x)/\sqrt{N_c}$ and
$(\vec{\pi}(x)-\vec{\pi}_0 (x)) \mapsto \vec{\pi}(x)/\sqrt{N_c}$ and 
carrying out the Gaussian integral with respect to $\sigma (x)$ 
and $\vec{\pi} (x)$, we derive $W$ as
\begin{eqnarray}
W[J_{\sigma}, \vec{J}_{\pi}] = I_0 
+ \frac{1}{2 N_c} \mathrm{Tr} \ln \mbit{I}^{(2)} 
+ O \left( \frac{1}{N_c^2} \right) \ , \label{functionW}
\end{eqnarray}
where Tr is the trace for mesonic degrees of freedom 
and the spacetime coordinate. 
Equation (\ref{functionW}) shows that the expansion (\ref{Iexpansion}) 
introduced above turns out to be the $1/N_c$ expansion. 
Inserting (\ref{functionW}) into (\ref{fields}), we get the relation 
between the classical fields and the corresponding fields 
$\varphi_{\sigma}(x)$ and $\vec{ \varphi}_{\pi}(x)$: 
\begin{eqnarray}
&& \varphi_{\sigma}(x) = \sigma_0 (x) + \frac{G}{2N_c}
  \frac{\delta}{\delta J_{\sigma}(x)} (\mathrm{Tr} \ln \mbit{I}^{(2)})
 = \sigma_0(x) + \frac{\sigma_1 (x)}{N_c} \ , \\
&& \vec{ \varphi}_{\pi}(x) = \pi_0 (x) + \frac{G}{2N_c}
  \frac{\delta}{\delta \vec{J}_{\pi}(x)} (\mathrm{Tr} \ln \mbit{I}^{(2)})
 = \vec{\pi}_0(x) + \frac{\vec{\pi}_1 (x)}{N_c} \ . 
\end{eqnarray}
The Legendre transformation of Eq.~(\ref{functionW}) leads to 
\begin{eqnarray}
 \mathcal{V}(\bar{\varphi}_{\sigma}, \vec{\bar{\varphi}}_{\pi}) 
  &=& \mathcal{V}_0 (\bar{\varphi}_{\sigma}, \vec{\bar{\varphi}}_{\pi}) 
+ \mathcal{V}_1 (\bar{\varphi}_{\sigma}, \vec{\bar{\varphi}}_{\pi}) \ , 
\label{effectivepotential}
\end{eqnarray}
\begin{eqnarray}
  \mathcal{V}_0 (\bar{\varphi}_{\sigma}, \vec{\bar{\varphi}}_{\pi})  
  &=& \frac{1}{2G} (\bar{\varphi}_{\sigma}^2 + \vec{\bar{\varphi}}_{\pi}^2)
  - \frac{1}{\beta V} 
  \ln \det (\gamma_{\mu} \partial_{\mu} + m_0 -\mu \gamma_4 + 
  \bar{\varphi}_{\sigma} 
  + i \gamma_5 \vec{\tau} \cdot \vec{\bar{\varphi}}_{\pi}) \ , \\
\mathcal{V}_1 (\bar{\varphi}_{\sigma}, \vec{\bar{\varphi}}_{\pi}) 
&=& \frac{1}{2 N_c \beta V} \mathrm{Tr} \ln 
  \mbit{I}^{(2)} (\bar{\varphi}_{\sigma}, \vec{\bar{\varphi}}_{\pi}) 
  + O \left(\frac{1}{N_c^2} \right) \ .
\end{eqnarray}
The leading term $\mathcal{V}_0$ corresponds 
to the contribution of the mean field approximation 
and the the next-to-leading contribution $\mathcal{V}_1$ to the contribution of meson loops of $1/N_c$ order.

Actual calculation of the effective potential~(\ref{effectivepotential}) 
is done in the momentum representation. 
The mean field part $\mathcal{V}_0$ is represented as 
\begin{eqnarray}
 \mathcal{V}_0 &=&
\frac{1}{2G} ((M-m_0)^2 + \vec{\bar{\varphi}}_{\pi}^2) \nonumber \\
&-&2N_f \int \frac{d^3 p}{(2 \pi)^3} 
\left[E_{\mbit{p}} + \frac{1}{\beta} 
\ln (1+\mathrm{e}^{-\beta (E_{\mbit{p}}+\mu)})
+ \frac{1}{\beta} 
\ln (1+\mathrm{e}^{-\beta (E_{\mbit{p}}-\mu)}) \right] \ ,
\label{meanfieldpotential}
\end{eqnarray}
where $E_{\mbit{p}} = (\mbit{p}^2+M^2+\vec{\bar{\varphi}}_{\pi}^2)^{1/2}$ and 
$M = m_0 + \bar{\varphi}_{\sigma}$. 
The mesonic loop correction part $\mathcal{V}_1$ becomes 
\begin{eqnarray}
 \mathcal{V}_1 = \frac{1}{2 N_c} [A_{\sigma}(T, \mu)+ 3 A_{\pi}(T, \mu)]
  \ ,
\label{v1}
\end{eqnarray}
\begin{eqnarray}
 A_{\hat{a}} = \frac{1}{\beta} \sum_{n = - \infty}^{\infty} 
\int \frac{d^3 p}{(2 \pi)^3} \ln \left(\frac{1}{G} - 
I_1 (i \omega_n^b, \mbit{p}) 
+ ((i \omega_n^b)^2 - \mbit{p}^2 -\varepsilon_{\hat{a}}^2) I_2(i
\omega_n^b, \mbit{p})   \right) \ ,
\label{1looppotential}
\end{eqnarray}
where
\begin{eqnarray}
 I_1(i \omega_n^b, \mbit{p}) &=& 2 N_f \bigg[ 
\int \frac{d^3 q}{(2 \pi)^3} \frac{1}{2 E_{\mbit{q}}}
\left[1-f(E_{\mbit{q}}+\mu)-f(E_{\mbit{q}}-\mu)  \right] 
\nonumber \\
&&+\int \frac{d^3 q}{(2 \pi)^3} \frac{1}{2 E_{\mbit{p}+\mbit{q}}}
\left[1-f(E_{\mbit{p}+\mbit{q}}+\mu)-f(E_{\mbit{p}+\mbit{q}}-\mu)  \right]
\bigg] \ ,  
\label{I1}
\end{eqnarray}
\begin{eqnarray}
&&\mathrm{Re} I_2(i \omega_n^b, \mbit{p}) = -\frac{N_f}{4} \int
\frac{d^3 q}{(2 \pi)^3} \frac{1}{E_{\mbit{p}+\mbit{q}} E_{\mbit{q}}} \nonumber \\
&&\times \Bigg[
\frac{E_{\mbit{p}+\mbit{q}}-E_{\mbit{q}}}
{{\omega_n^b}^2 +(E_{\mbit{p}+\mbit{q}}-E_{\mbit{q}})^2 }
(f(E_{\mbit{p}+\mbit{q}}+\mu) + f(E_{\mbit{p}+\mbit{q}}-\mu) 
-f(E_{\mbit{q}}+\mu) - f(E_{\mbit{q}}-\mu) ) \nonumber \\
&&+ \frac{E_{\mbit{p}+\mbit{q}}+E_{\mbit{q}}}
{{\omega_n^b}^2 +(E_{\mbit{p}+\mbit{q}}+E_{\mbit{q}})^2 }
(2-f(E_{\mbit{p}+\mbit{q}}+\mu) - f(E_{\mbit{p}+\mbit{q}}-\mu) 
-f(E_{\mbit{q}}+\mu) - f(E_{\mbit{q}}-\mu) )
\Bigg] \ ,
\label{ReI2}
\end{eqnarray}
\begin{eqnarray}
 \mathrm{Im} I_2(i \omega_n^b, \mbit{p}) &=& N_f \int
\frac{d^3 q}{(2 \pi)^3} \frac{\omega_n^b}
{({\omega_n^b}^2 + E_{\mbit{p}+\mbit{q}}^2 + E_{\mbit{q}}^2)^2
-4 E_{\mbit{p}+\mbit{q}}^2 E_{\mbit{q}}^2}
\nonumber \\
&& \times
(f(E_{\mbit{q}}+\mu) - f(E_{\mbit{q}}-\mu) 
-f(E_{\mbit{p}+\mbit{q}}+\mu) + f(E_{\mbit{p}+\mbit{q}}-\mu) ) \ , 
\label{ImI2}
\end{eqnarray}
with $\omega_n^b=2\pi n /\beta$ 
and $f(E) \equiv 1/(\mathrm{e}^{\beta E} + 1)$. 
In the limit of zero temperature, these equations are reduced to simpler forms:
\begin{eqnarray}
\mathcal{V}_0 &\stackrel{T \to 0}{\longrightarrow}&
\frac{1}{2G} ((M-m_0)^2 + \vec{\bar{\varphi}}_{\pi}^2)
-2N_f \int \frac{d^3 p}{(2 \pi)^3} 
\left[\mu+(E_{\mbit{p}}-\mu) \theta (E_{\mbit{p}}-\mu) \right] \ ,
\end{eqnarray}
\begin{eqnarray}
 A_{\hat{a}} &\stackrel{T \to 0}{\longrightarrow}&
\int \frac{d \omega}{2 \pi}
\int \frac{d^3 p}{(2 \pi)^3} \ln \left(\frac{1}{G} - 
I_1 (i \omega, \mbit{p}) 
+ ((i \omega)^2 - \mbit{p}^2 -\varepsilon_{\hat{a}}^2) I_2(i
\omega, \mbit{p})   \right) \ ,
\end{eqnarray}
\begin{eqnarray}
I_1(i \omega, \mbit{p}) 
&=& 
2 N_f \bigg[ 
\int \frac{d^3 q}{(2 \pi)^3} \frac{1}{2 E_{\mbit{q}}} 
\theta (E_{\mbit{q}}-\mu)
+\int \frac{d^3 q}{(2 \pi)^3} \frac{1}{2 E_{\mbit{p}+\mbit{q}}}
\theta (E_{\mbit{p}+\mbit{q}}-\mu) 
\bigg] \ ,
\end{eqnarray}
\begin{eqnarray}
 \mathrm{Re} I_2(i \omega, \mbit{p}) 
&=& 
-\frac{N_f}{4} \int
\frac{d^3 q}{(2 \pi)^3} \frac{1}{E_{\mbit{p}+\mbit{q}} E_{\mbit{q}}}
\nonumber \\
&& \times \Bigg[
\frac{E_{\mbit{p}+\mbit{q}}-E_{\mbit{q}}}
{{\omega}^2 +(E_{\mbit{p}+\mbit{q}}-E_{\mbit{q}})^2 }
(\theta (E_{\mbit{q}}-\mu) - \theta (E_{\mbit{p}+\mbit{q}}-\mu))
\nonumber \\
&& 
+ \frac{E_{\mbit{p}+\mbit{q}}+E_{\mbit{q}}}
{{\omega}^2 +(E_{\mbit{p}+\mbit{q}}+E_{\mbit{q}})^2 }
(\theta (E_{\mbit{q}}-\mu) + \theta (E_{\mbit{p}+\mbit{q}}-\mu))
\bigg] \ ,
\end{eqnarray}
\begin{eqnarray}
&&\mathrm{Im} I_2(i \omega, \mbit{p}) \nonumber \\ 
&=& 
N_f \int
\frac{d^3 q}{(2 \pi)^3} \frac{\omega}
{({\omega}^2 + E_{\mbit{p}+\mbit{q}}^2 + E_{\mbit{q}}^2)^2
-4 E_{\mbit{p}+\mbit{q}}^2 E_{\mbit{q}}^2}
(\theta (E_{\mbit{q}}-\mu) - \theta (E_{\mbit{p}+\mbit{q}}-\mu)) \ ,
\end{eqnarray}
where note that $\omega$ becomes a continuous valuable.

In the present analysis, 
we adopt the parameter set of Hatsuda and Kunihiro~\cite{HK2} 
which are determined 
so as to reproduce $F_{\pi} = 93$ MeV and $m_{\pi} = 138$ MeV at $T=\mu=0$;
the resultant values are 
$m_0=5.5$ MeV, the four-Fermi coupling constant $G= 32.976$ GeV$^{-2}$ and 
the ultra-violet divergence cutoff $\Lambda = 631$ MeV.

\begin{figure}[htbp]
\begin{center}
 \includegraphics[width=7.5cm]{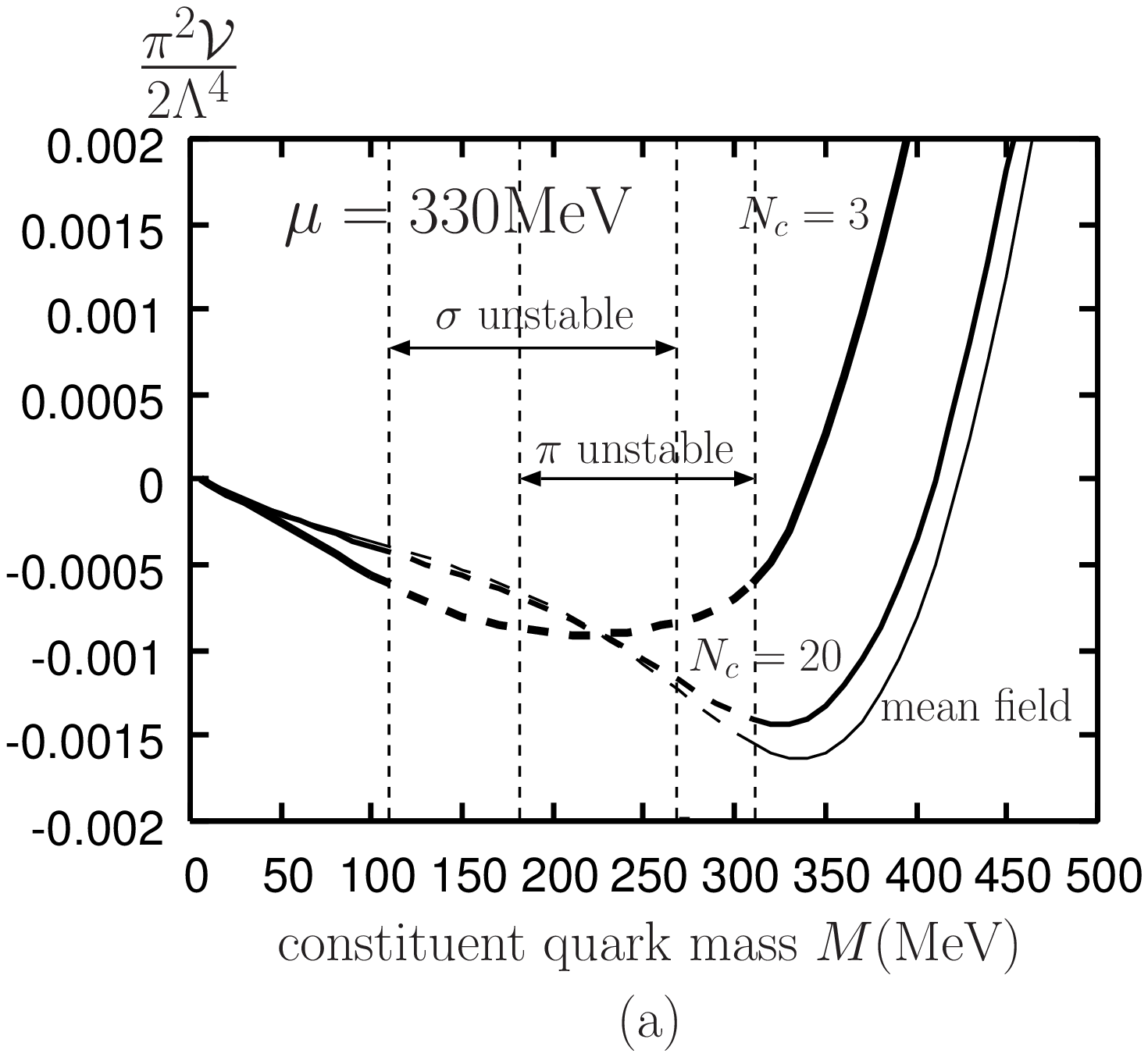} 
 \includegraphics[width=7.5cm]{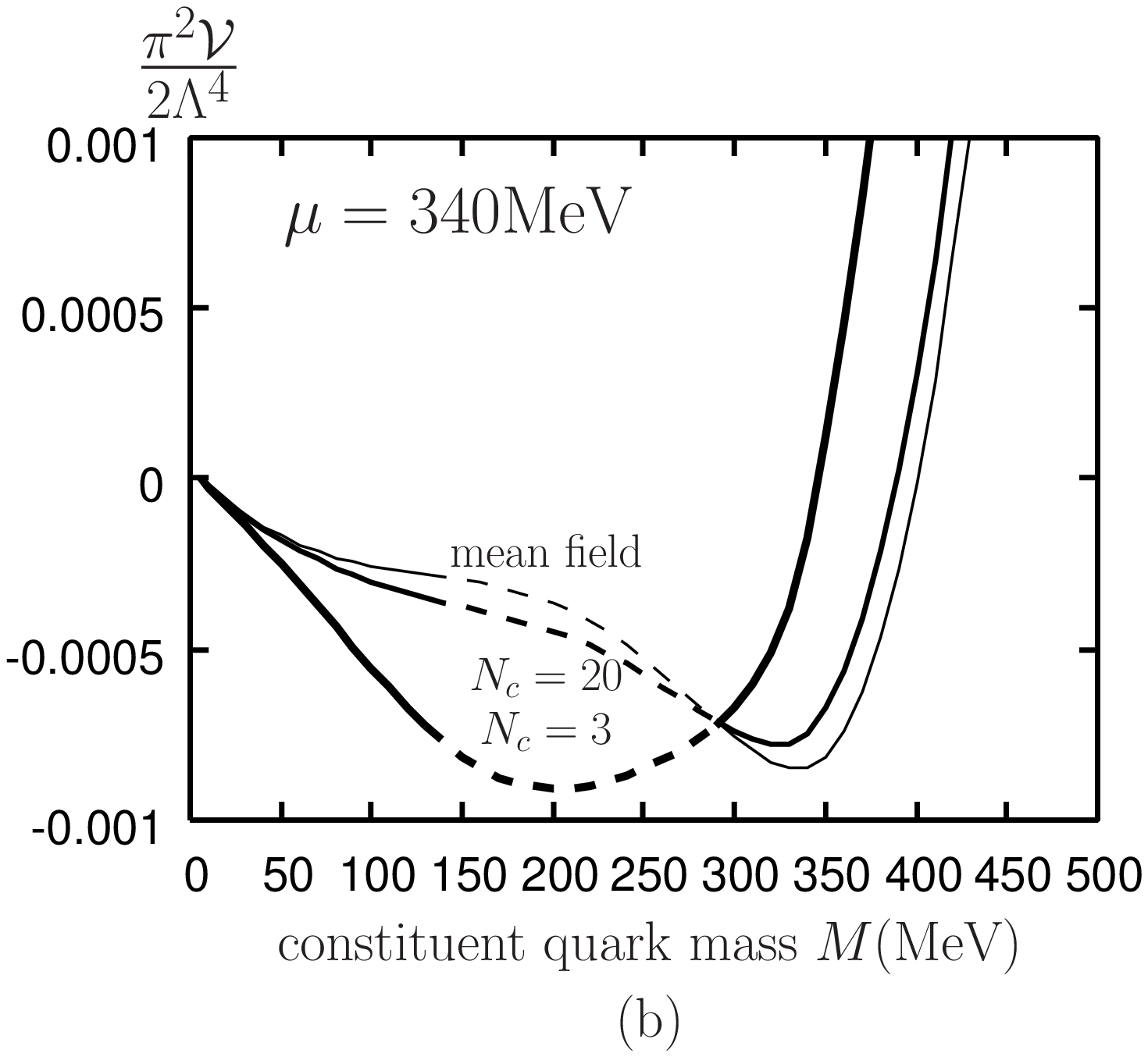}
 \includegraphics[width=7.5cm]{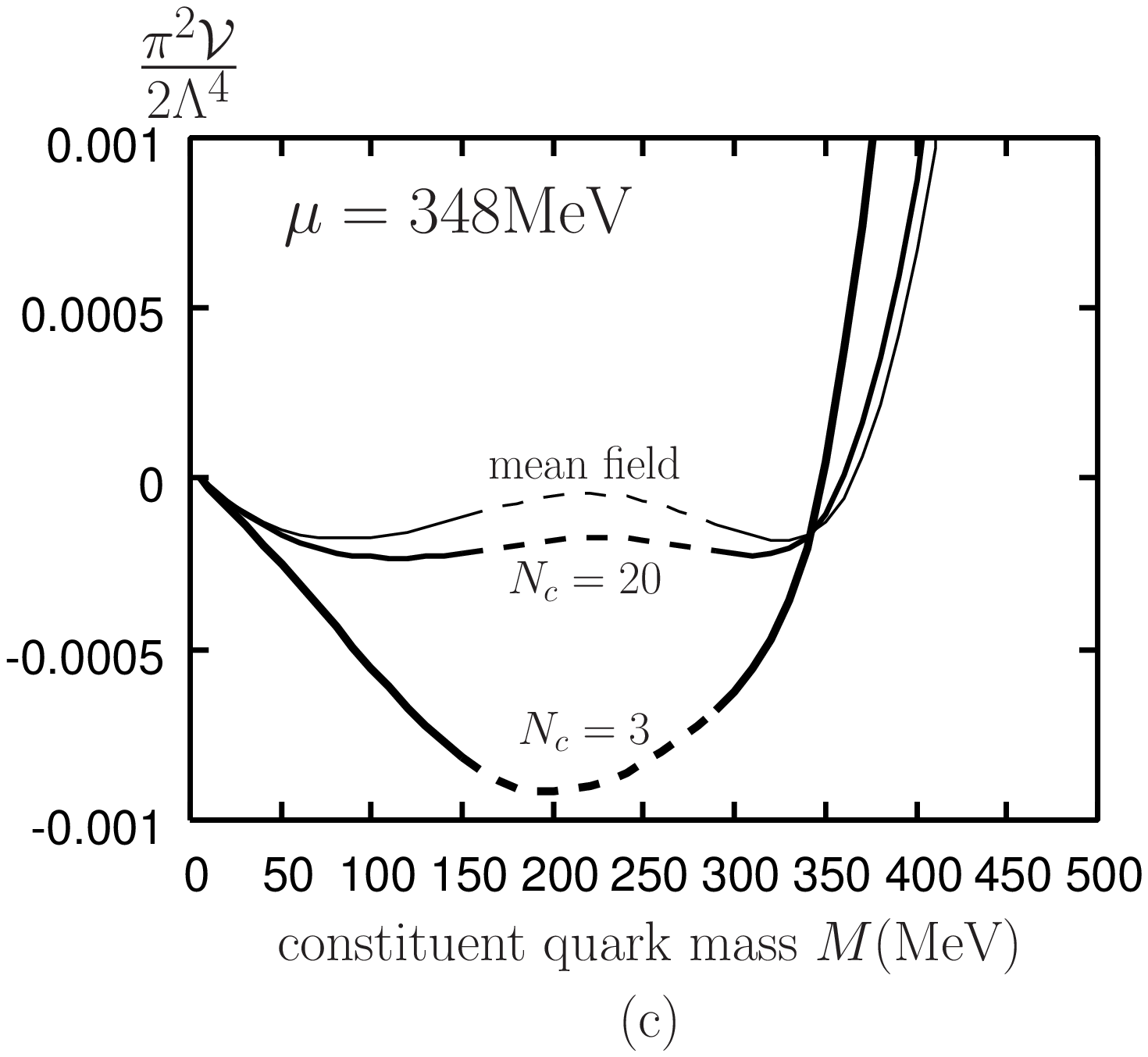} 
 \includegraphics[width=7.5cm]{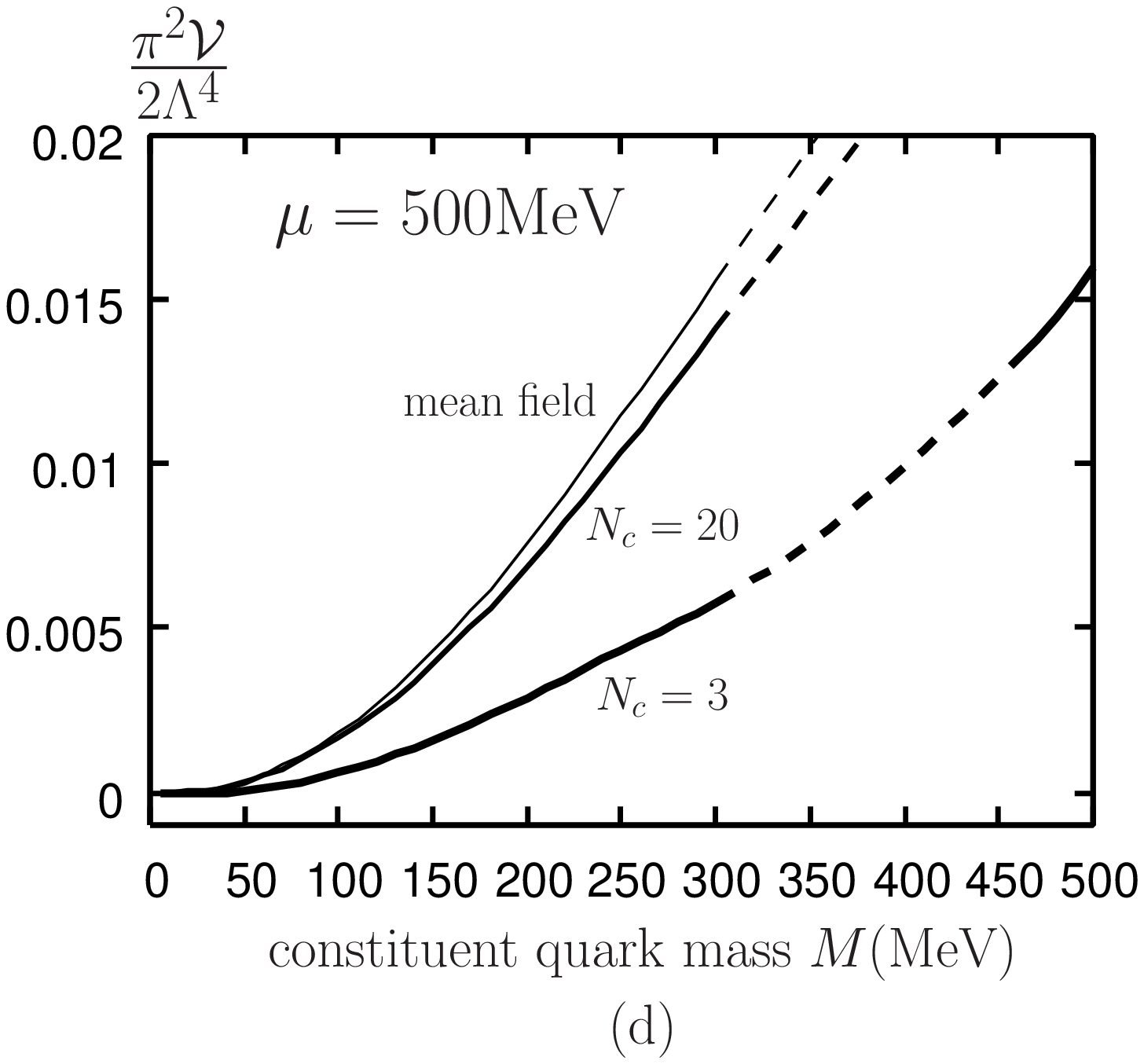}
\end{center}
\caption{Non-dimensionalized effective potential as a function of the 
constituent quark mass. Cases of four different chemical potentials are 
presented. Thickness of curves decreases as the number of color increases 
as $3\rightarrow20\rightarrow\infty$. 
Dashed curves indicate regions unstable 
with respect to the $\sigma$ and/or $\pi$ directions. 
For $\mu=330$~MeV, regions unstable in the $\sigma$ ($\pi$) direction are 
denoted by ``$\sigma$ unstable" (``$\pi$ unstable").
For $\mu > 330$~MeV, all regions denoted by dashed curves are unstable only 
in the $\sigma$ direction. 
}
\label{fig1}
\end{figure}

Figure~\ref{fig1} represents the effective potential 
at $\mu = 330, 340, 348$ and $500$~MeV. 
When $\mu = 330$ MeV, there appears a minimum 
around $M=340$~MeV at the mean field level.
The stability of the mean field solution can be investigated by 
the curvature of $\mathcal{V}_0$ in the $\bar{\varphi}_{\sigma}$ and 
$\bar{\varphi}_{\pi}$ directions. 
In the region denoted by ``$\sigma$ unstable", 
the curvature is negative and then unstable 
in the $\bar{\varphi}_{\sigma}$ direction. 
Similarly, in the region denoted by ``$\pi$ unstable", 
the curvature is negative in the $\bar{\varphi}_{\pi}$ direction. 
Fortunately, the minimum around $M=340$~MeV is out of the unstable regions. 
This property is held for other values of $\mu$; 
three examples are shown in other panels of Fig.~\ref{fig1}. 
As another interesting point, any $\pi$ unstable region does not appear 
for $\mu > 330$ MeV. Thus, the mean field solution at the minimum point is 
stable at any $\mu$ for the case of the present parameter set.

The meson loop corrections in the unstable regions 
do not make sense, since mesons considered there are tachyonic. 
Actually, the Gaussian integral in Eq.(\ref{auxiliarypartition}) breaks down 
for tachyonic mesons. Note that 
dashed curves in the unstable regions are just a guide of eyes. 
In the $N_c=3$ case, we can not see where is a minimum, 
since it is somewhere in the unstable regions. 
So we take a somewhat larger $N_c$, or 20, in which 
a minimum is still out of the unstable regions 
even after the inclusion of the correction. 
We then look into the effect of the next-to-leading order correction by 
comparing the mean field solution (the $N_c\rightarrow\infty$ case) 
and the finite $N_c$ ($=20$) case.  

 First we discuss the result of the mean field approximation.
Studies in the mean field approximation level have already been done in 
Refs.~\cite{BR,AY,Sca,Fuj}. At $\mu=$ 340 MeV (Fig.~\ref{fig1}(b)) the 
minimum is still located around $M=$ 340 MeV; this means that chiral 
symmetry is broken. At $\mu=$ 348 MeV (Fig.~\ref{fig1}(c)) two minima 
degenerate; in other words this is the first order phase transition 
point. At higher $\mu$ (Fig.~\ref{fig1}(d)) chiral symmetry is restored 
to some extent. Here a comment is in order. In this case $M$ at the minimum 
is still around 50 MeV, and it decreases gradually as $\mu$ increases; 
in other words we see a crossover to the restored phase. This situation was 
seen also in Refs.~\cite{AY,Sca} although not discussed. 

\begin{figure}[htbp]
\begin{center}
 \includegraphics[width=9cm]{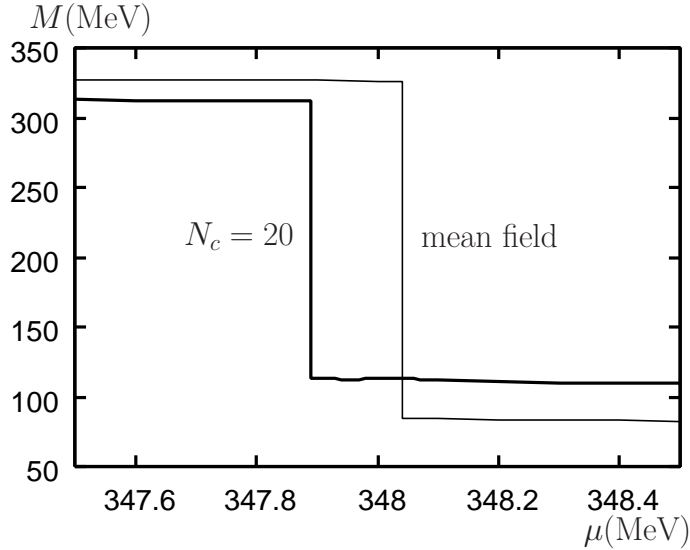} 
\end{center}
\caption{The value of the constituent quark mass at which the effective potential 
becomes minimum as a function of the chemical potential. The mean field and 
the $N_c=20$ cases are presented.}
\label{fig2}
\end{figure}

 Now we consider the next-to-leading order correction due to a finite 
$N_c$. In the case of $N_c=20$ the results are similar to the mean field 
case but the correction weakens the transition, in other words makes the 
jump in $M$ small and shifts the critical $\mu$ to a lower value. 
These are clearly shown in Fig.~\ref{fig2}. 

To summarize, we have studied semi-quantitatively the effects of the mesonic, 
i.e., the next-to-leading order in the $1/N_c$ expansion, correction in the 
Nambu--Jona-Lasinio model on the high density chiral phase transition based 
on the auxiliary field method. The finite $N_c$ correction weakens the 
first order phase transition and shifts the critical chemical potential to a 
lower value. At $N_c=3$, however, we can not see the minimum because of the 
instability of the mean field effective potential to the direction of the 
$\sigma$ and/or $\pi$ classical fields. For the mean field calculation, 
we have pointed out explicitly that chiral symmetry is not completely restored 
immediately at the first order transition; after the first order transition 
the constituent quark mass decreases gradually --- this may be called the 
two step restoration. 

 We still have a lot to do: First of all we have to redetermine 
the parameter set, 
the four-Fermi coupling constant $G$ and the ultra-violet cutoff $\Lambda$ 
in the next-to-leading order. Next we have to study the phase diagram 
at finite $T$ and $\mu$ to 
find the location of the critical end point. Further, studies of the three 
flavor model and the color superconductivity phase are to be done.

\end{document}